\documentclass[twoside,english,final,5p]{elsarticle}
\usepackage{amssymb}
\usepackage{bbm}
\usepackage{amssymb,graphicx}
\usepackage[intlimits]{amsmath}
\usepackage[small]{subfigure}
\usepackage{color}
\usepackage{array}
\usepackage{multirow}

\usepackage[colorlinks=true]{hyperref}
\hypersetup{
 linkcolor=blue,citecolor=blue,urlcolor=black}
 
\makeatletter
\@ifundefined{date}{}{\date{}}

\journal{Physics Letters B}

\makeatother

\usepackage{babel}
\begin{document}

\begin{frontmatter}{}

\title{Boundary Integrability from the Fuzzy Three Sphere}

\author[bdt,wig]{T.\ Gombor}
\author[tri]{A.\ Holguin}

\affiliation[bdt]{organization={Department of Theoretical Physics, Eötvös Loránd University },
           addressline={Pázmány Péter sétány 1A}, 
            city={Budapest},
            postcode={1117}, 
           country={Hungary}}
            
\affiliation[wig]{organization={HUN-REN Wigner Research Centre for Physics},
            addressline={Konkoly-Thege Miklós út 29-33.}, 
         city={Budapest},
        postcode={1121}, 
         country={Hungary}}

\affiliation[tri]{organization={School of Mathematics, Trinity College Dublin},
           addressline={17 Westland Row}, 
           city={Dublin},
            postcode={Dublin 2}, 
            country={Ireland}}
            
\begin{abstract}
We consider $\mathfrak{so}_4$ invariant matrix product states (MPS) in the $\mathfrak{so}_6$  symmetric integrable spin chain and prove their integrability. These MPS appear as fuzzy three-sphere solutions of matrix models with Yang-Mills-type interactions, and in particular they correspond to scalar defect sectors of $\mathcal{N}=4$ SYM. We find that the algebra formed by the fuzzy three-sphere generators naturally leads to a boundary reflection algebra and hence a solution to the boundary Yang-Baxter equation for every representation of the fuzzy three-sphere. This allows us to find closed formula for the overlaps of Bethe states of $\mathfrak{so}_6$ symmetric chains with the fuzzy three-sphere MPS for arbitrary bond dimensions. \end{abstract}

\end{frontmatter}{}

\section{Introduction}
Exact results for real time evolution of interacting quantum systems are of increasing importance due to the growing interest in simulations in quantum computers and in analog optical traps \cite{Hackerm_ller_2010, Schneider2012FermionicTA, PhysRevLett.122.090601, Bauer:2022hpo}.  On a practical level one would like to have exactly solvable quantum models which are simple enough to test simulations, yet rich enough to have non-trivial dynamics. Integrable models in 1+1 dimensions are of course ideal for such a task, however despite their inherent solvability, the computations of real time observables is complicated by the inherent complexity of their many body wavefunctions. Recent progress towards this goal has been achieved via the identification of certain integrable states for which overlaps with energy eigenstates take a simple and universal form \cite{Calabrese_2016}. By now there is a compelling picture relating these integrable quenches to integrability in the presence of boundaries in a rotated channel and the universal form of the overlap formulas have been established in very general settings \cite{Piroli:2017sei, Pozsgay:2018dzs, Gombor:2021uxz, Gombor:2021hmj, Gombor:2024iix}. 

One particularly interesting class of such states are those which arise from matrix product states (MPS). Many of the known examples of integrable MPS come from studies of defect set-ups in the context of the AdS/CFT correspondence \cite{deLeeuw:2015hxa, Buhl-Mortensen:2015gfd, deLeeuw:2016ofj, Kristjansen:2024map, deLeeuw:2024qki, Holguin:2025bfe, Chalabi:2025nbg, Kristjansen:2024dnm}. There the states have a natural interpretation as boundary states for the dual string description of the large $N$ gauge theory, and the overlaps of the integrable MPS encode rich information about non-perturbative physics in the string theory. One of the prospects of said program is the determination of finite coupling OPE data in superconformal theories such as $\mathcal{N}=4$ SYM and ABJM \cite{Jiang:2019zig}. 

The duality between integrable spin systems describing the dynamics of single trace operators in the gauge theory and strings moving in anti-de-Sitter spaces opens an avenue for generating large classes of boundary states for spin systems inspired by known results in string theory. A common effect in string theories is that boundary conditions for open strings carry non-commutative data which is often geometric. This is well understood for instance in Yang-Mills matrix models in which brane configurations are often described by non commutative geometries \cite{Connes:1997cr}. Noncommutative geometries such as fuzzy spheres of various dimensions generically appear as classical configurations of matrix theories \cite{Madore:1991bw, Kabat:1997im,  Hoppe:2002km}, such as conformal defect set-ups of $\mathcal{N}=4$ SYM. These configurations can often be associated with integrable MPS making them interesting independently of their stringy origins. This motivated the conjecture that in fact all fuzzy spheres can be associated with an integrable MPS  which led to the identification of novel integrable MPS based on odd dimension fuzzy spheres \cite{deLeeuw:2024qki}.

In this letter we establish the integrability of a class of $\mathfrak{so}_4$ symmetric MPS based on the fuzzy three-sphere for $\mathfrak{so}_n$ symmetric integrable chains. For applications to $\mathcal{N}=4$ SYM we will keep in mind the case of $n=6$. Such MPS is the simplest integrable quench based on odd dimensional fuzzy spheres  and it demonstrates the inherent difference of $\mathfrak{so}_m$ invariant MPS for odd $m$ relative to those of even $m$.  The algebra generated by the fuzzy $S^3$ MPS is enough to guarantee the existence of a K-matrix giving a representation of the boundary reflection algebra for every rank for which the MPS exists. With this information we establish an overlap formula for the $\mathfrak{so}_6$ symmetric chain. 

We begin in section \ref{fuzzy s3} by presenting the fuzzy $S^3$ algebra. In section \ref{integrability so(6)} we review the integrability framework for computing overlaps of integrable MPS in the $SO(6)$ symmetric spin chain. In  section \ref{K matrix} we present the matrix for the fuzzy $S^3$ MPS and the corresponding Cartan generators of the reflection algebra. With this information we present the overlap formula in section \ref{overlap formula}. Finally, in section \ref{conclusion}, we conclude.

\section{Fuzzy $S^3$ Algebra}\label{fuzzy s3}
We begin by presenting the fuzzy $S^3$ algebra. An $SO(4)$ invariant construction of the fuzzy three-sphere was presented in \cite{Ramgoolam:2001zx}. This method can be used to construct matrices satisfying the matrix equation $\sum_{i=1}^4 X_i^2=r_n\mathbb{I}$, where the rank of the matrices is given by $r_n=\frac{(n+1)(n+3)}{2}$. The meaning of $n$ will become clear momentarily. Instead of reviewing their explicit construction we instead summarize the results found in \cite{Ramgoolam:2002wb} which presents an algebra generated from the $X_i$. The algebra presented by Ramgoolam can be summarized by the following set of relations:
\begin{align}
\left\{ X_{j},\mathcal{U}\right\}  & =\left[F_{j,k},\mathcal{U}\right]=0,\quad\mathcal{U}^{2}=1,\\
\sum_{k=1}^{4}X_{k}^{2} & =\frac{(n+1)(n+3)}{2}\mathbb{I},\label{eq:Xsquare}\\
\left[X_{j},X_{k}\right] & =F_{jk}+(n+2)F_{jk}^{-},\\
\left[F_{jk},X_{l}\right] & =\delta_{kl}X_{j}-\delta_{jl}X_{k}.\\
\left[F_{ij},F_{kl}\right] & =\delta_{jk}F_{il}-\delta_{il}F_{kj}-\delta_{ik}F_{jl}+\delta_{jl}F_{ki}.
\end{align}
The  $F_{ij}$ act as $SO(4)$ generators when acting the $X_i$, while $\mathcal{U}$ acts as a chirality operator. The $F^{-}_{ij}$ are not independent and they are defined by
\begin{equation}
F_{ij}^{-}=\frac{1}{2}\sum_{kl}\epsilon_{ijkl}F_{kl}\mathcal{U}.
\end{equation}
Since the sum of the square of the coordinate matrices $X_i$ is central in the algebra,
\begin{equation}
\left[\sum_{k=1}^{4}X_{k}^{2},X_{j}\right]=0,
\end{equation}
we have some additional relations that are useful for determining the $K$-matrix:
\begin{equation}
\sum_{j=1}^{4}X_{j}F_{jk}=\frac{3}{2}X_{k},
\end{equation}
\begin{equation}
\sum_{j=1}^{4}X_{j}F_{jk}^{-}=\left(\frac{n+2}{2}\right)X_{k}.
\end{equation}
Representations of this algebra act on a reducible representation of $SO(4)$  $\mathcal{R}_+ \oplus \mathcal{R}_-= \left(\frac{n+1}{4}, \frac{n-1}{4}\right)\oplus \left(\frac{n-1}{4}, \frac{n+1}{4}\right)$ on which the generator $\mathcal{U}$ acts diagonally with eigenvalues $\pm1$. The matrices $X_i$ are maps between the irreducible components $\mathcal{R}_+$ and $\mathcal{R}_-$ while the generators $F_{ij}$ are symmetry generators. 

\section{Integrable SO(6) spin chains and their overlaps} \label{integrability so(6)}

The planar spectral problem of the $\mathcal{N}=4$ SYM is solved
by an integrable super spin chain \cite{Beisert:2005fw}. In particular, at the one-loop
level the single trace operators in the scalar sector can be described
by state of the $SO(6)$ symmetric spin chain. The Hamiltonian is \cite{Minahan:2002ve}
\begin{equation}
H=\sum_{j=1}^{J}1-P_{j,j+1}+\frac{1}{2}K_{j,j+1},
\end{equation}
where $P$ and the $K$ are the permutation and trace operators acting
on the tensor product of two copies of 6 dimensional vector space
\begin{equation}
P=\sum_{i,j=1}^{6}e_{i,j}\otimes e_{j,i},\quad K=\sum_{i,j=1}^{6}e_{i,j}\otimes e_{i,j},
\end{equation}
where $e_{i,j}$ are the unit $6\times6$ matrices. The $J$ is the
length of the spin chain. 

The Hamiltonian is integrable which means there exist a transfer matrix
which generates infinte series of conserved charges. The transfer
matrix is defined from the monodromy matrix
\begin{equation}
T_{0}(u)=R_{0,J}(u-1)\dots R_{0,1}(u-1),
\end{equation}
where $R(u)$ is the $SO(6)$ symmetric $R$-matrix
\begin{equation}
R_{0,j}(u)=1+\frac{1}{u}P_{0,j}-\frac{1}{u+2}K_{0,j}.
\end{equation}
These operators are written in the real basis of the scalar fields.
We also need the form in the complex basis
\begin{equation}
\begin{split}|-3\rangle & =Z=\frac{1}{\sqrt{2}}(\phi_{1}+i\phi_{6}),\quad|3\rangle=\bar{Z}=\frac{1}{\sqrt{2}}(\phi_{1}-i\phi_{6}),\\
|-2\rangle & =Y=\frac{1}{\sqrt{2}}(\phi_{2}+i\phi_{5}),\quad|2\rangle=\bar{Y}=\frac{1}{\sqrt{2}}(\phi_{2}-i\phi_{5}),\\
|-1\rangle & =X=\frac{1}{\sqrt{2}}(\phi_{3}+i\phi_{4}),\quad|1\rangle=\bar{X}=\frac{1}{\sqrt{2}}(\phi_{3}-i\phi_{4}).
\end{split}
\end{equation}
In this new basis the permutation has the same form but the trace
operator reads as
\begin{equation}
K=\sum_{i,j=-3}^{3}e_{i,j}\otimes e_{-i,-j},
\end{equation}
where sum goes through the indexes $i,j=-3,-2,-1,1,2,3$. The dictionary
between Bethe eigenstates $|\mathbf{\ensuremath{u}},\mathbf{\ensuremath{v}},\mathbf{\ensuremath{w}}\rangle$
and operators built from complex fields can be found for instance
in \citep{Minahan:2002ve} \footnote{The three sets of Bethe roots are associated with nodes of the $SO(6)$
Dynkin diagram. The one associated with the middle node (momenta carrying
excitations) is denoted by $\mathbf{\ensuremath{u}}$. The other two
nodes correspond to the sets $\mathbf{v}$ and $\mathbf{w}$, respectively.}.

Let us define the MPS without the trace
\begin{equation}
\langle\psi_{\alpha,\beta}|=\sum_{i_{1},\dots,i_{J}}(\omega_{i_{j}}\dots\omega_{i_{1}})_{\alpha,\beta}\langle i_{1},\dots,i_{j}|,
\end{equation}
where $\alpha,\beta$ are indexes in the bond space. The integrable
MPS is defined with the $KT$-relation
\begin{equation}
\sum_{k,\gamma}K_{i,k}^{\alpha,\gamma}(u)\langle\psi_{\gamma,\beta}|T_{k,j}(u)=\sum_{k,\gamma}\langle\psi_{\alpha,\gamma}|T_{i,k}(-u)K_{k,j}^{\gamma,\beta}(u),
\end{equation}
where the $\alpha,\beta,\gamma$ and $i,j,k$ are indexes in the bond
and the six dimensional space, respectively. Defining operators in
the bond space as $\mathbf{K}_{i,j}=\sum_{\alpha,\beta}K_{i,j}^{\alpha,\beta}(u)e_{\alpha,\beta}^{B}$
and $\langle\Psi|=\sum_{\alpha,\beta}\langle\psi_{\alpha,\beta}|\otimes e_{\alpha,\beta}^{B}$
we obtain more compact form of the $KT$-equation
\begin{equation}
\mathbf{K}_{0}(u)\langle\Psi|T_{0}(u)=\langle\Psi|T_{0}(-u)\mathbf{K}_{0}(u).
\end{equation}
The MPS overlap formula can be expressed from the components of the
$K$-matrix in the complex basis \citep{Gombor:2024iix,Gombor:2025wvu}.
Since our MPS has $SO(4)\times SO(2)$ symmetry, the Bethe vectors
with non-vanish overlaps have achiral pair structure \citep{Gombor:2020kgu}
for which $\mathbf{v}=-\mathbf{w}$ and $\mathbf{u}=-\mathbf{u}$.
The overlaps can be obtained in the following way. At first, we define
the nested $K$-matrix 
\begin{equation}
\mathbf{K}_{i,j}^{(2)}(u)=\mathbf{K}_{i,j}(u)-\mathbf{K}_{i,-3}(u)\mathbf{K}_{3,-3}^{-1}(u)\mathbf{K}_{3,j}(u),
\end{equation}
for $i,j=-2,-1,1,2$. After that, we calculate the $G$ operators
as
\begin{equation}
\begin{split}\mathbf{G}^{(1)}(u) & =\mathbf{K}_{3,-3}(u),\\
\mathbf{G}^{(2)}(u) & =\mathbf{K}_{2,-2}^{(2)}(u),\\
\mathbf{G}^{(3)}(u) & =\mathbf{K}_{1,1}^{(2)}(u)-\mathbf{K}_{1,-2}^{(2)}(u)\left[\mathbf{K}_{2,-2}^{(2)}(u)\right]^{-1}\mathbf{K}_{1,-2}^{(2)}(u).
\end{split}
\end{equation}
and the $\mathbf{B}$-operator as
\begin{equation}
\mathbf{B}=\langle\Psi|0\rangle.
\end{equation}
These operators commute, i.e.,
\begin{equation}
\left[\mathbf{B},\mathbf{G}^{(s)}(u)\right]=\left[\mathbf{G}^{(s)}(u),\mathbf{G}^{(r)}(v)\right]=0.
\end{equation}
Next step is to define the $F$-operators
\begin{equation}
\begin{split}\mathbf{F}^{(1)}(u) & =\left[\mathbf{G}^{(1)}(u)\right]^{-1}\mathbf{G}^{(2)}(u),\\
\mathbf{F}^{(2)}(u) & =\left[\mathbf{G}^{(2)}(u)\right]^{-1}\mathbf{G}^{(3)}(u).
\end{split}
\end{equation}
We can diagonalize these operators simultaneously
\begin{equation}
\begin{split}\mathbf{B} & =\mathbf{A}\mathrm{diag}(\beta_{1},\dots,\beta_{d})\mathbf{A}^{-1},\\
\mathbf{F}^{(s)} & =\mathbf{A}\mathrm{diag}(\mathcal{F}_{1}^{(s)},\dots,\mathcal{F}_{d}^{(s)})\mathbf{A}^{-1}.
\end{split}
\end{equation}
The MPS overlap is given by
\begin{multline}
\frac{\langle\mathrm{MPS}|\mathbf{u},\mathbf{v},\mathbf{w}\rangle}{\sqrt{\langle\mathbf{u},\mathbf{v},\mathbf{w}|\mathbf{u},\mathbf{v},\mathbf{w}\rangle}}=\\
\sum_{\ell=1}^{d}\beta_{\ell}\prod_{j=1}^{n_{u}/2}\tilde{\mathcal{F}}_{\ell}^{(1)}(u_{j})\prod_{j=1}^{n_{v}}\tilde{\mathcal{F}}_{\ell}^{(2)}(v_{j})\sqrt{\frac{\det G_{+}}{\det G_{-}}},\label{eq:genOV}
\end{multline}
where
\begin{equation}
\begin{split}\tilde{\mathcal{F}}_{\ell}^{(1)}(u) & =\mathcal{F}_{\ell}^{(1)}(iu+1)\sqrt{\frac{u^{2}}{u^{2}+1/4}},\\
\tilde{\mathcal{F}}_{\ell}^{(2)}(u) & =\mathcal{F}_{\ell}^{(2)}(iu)\frac{u}{u+i/2}.
\end{split}
\end{equation}
The Gaudin determinants with achiral pair
structure enters in this expression. Explicit expressions for them in our conventions can be
found in ref. \citep{Jiang:2019xdz,Gombor:2020kgu,Gombor:2021hmj,Gombor:2024iix}

\section{Fuzzy sphere K-matrix and the corresponding Cartan operators}\label{K matrix}

When the MPS is built from the operators
\begin{equation}
\omega_{j}=\begin{cases}
X_{j}, & j=1,2,3,4,\\
0 & j=5,6,
\end{cases}
\end{equation}
the $K$-matrix in the real basis can be expressed as
\begin{equation}
K_{i,j}=\begin{cases}
c_{1}\delta_{i,j}+X_{i}X_{j}+c_{2}X_{j}X_{i}+c_{3}F_{i,j}^{-}, & i,j=1,\dots,4,\\
c_{4}\delta_{i,j}, & i,j=5,6,\\
0, & \text{otherwise},
\end{cases}
\end{equation}
where
\begin{equation}
\begin{split}c_{1} & =\frac{u-1}{u-2}\frac{(4u^{2}-8u+n^{2}+4n+7)}{4(2u-1)},\\
c_{2} & =-\frac{u-1}{u-2},\\
c_{3} & =-\frac{2(n+2)(u-1)^{2}}{(u-2)(2u-1)},\\
c_{4} & =-\frac{u-1}{u-2}\frac{(2u-n-3)(2u+n+1)}{4(2u-1)}.
\end{split}
\end{equation}
The $F$-operators can be expressed as with the Casimir operators of
the subalgebras $\mathfrak{so}_{2}\subset\mathfrak{so}_{3}\subset\mathfrak{so}_{4}$:
\begin{equation}
\mathcal{C}_{k}=\frac{1}{2}\sum_{i,j=5-k}^{4}F_{i,j}^{2}.
\end{equation}
which generate commuting subalgbra in $\mathfrak{so}_{4}$. The $F$-operators
can be expressed with the Casimirs as 
\begin{equation}
\begin{split}\mathbf{F}^{(1)}(u) & =\frac{u(u-1)}{(u-\frac{1}{2})^{2}}\frac{(u-\frac{n+3}{2})(u+\frac{n+1}{2})\left[(u-\frac{1}{2})^{2}+\mathcal{C}_{2}\right]}{((u-1)^{2}-\frac{1}{4}+\mathcal{C}_{3})(u^{2}-\frac{1}{4}+\mathcal{C}_{3})},\\
\mathbf{F}^{(2)}(u) & =\frac{u-1/2}{u}\frac{u^{2}-\frac{1}{4}+\mathcal{C}_{3}}{(u+\frac{1}{2}-iF_{3,4})(u-\frac{1}{2}-iF_{3,4})}.
\end{split}
\end{equation}
where expressions in the denominator denote the invertation.

\section{Overlap formula}\label{overlap formula}

For the final overlap formula we only need to diagonalize the $F$-operators
which are expressed with the Casimirs. The Casimirs are diagonal in
the Gelfand-Tsetlin basis \cite{molev2002GT}. In the GT-basis, every basis vector
corresponds to a GT-pattern. For the representations $(\frac{n+1}{4},\frac{n-1}{4})$
or $(\frac{n-1}{4},\frac{n+1}{4})$, the possible GT-patterns are
\[
|k,l\rangle\Longleftrightarrow\begin{array}{ccc}
\frac{n}{2} &  & \pm\frac{1}{2}\\
 & k\\
 &  & l
\end{array}
\]
where $k,l$ are half integers between $\frac{1}{2}\leq k\leq\frac{n}{2}$
and $k-\leq l\leq k$. The Casimirs act diagonally on these basis
vectors 
\begin{align}
\mathcal{C}_{3}|k,l\rangle & =-k(k+1)|k,l\rangle,\nonumber \\
\mathcal{C}_{2}|k,l\rangle & =-l^{2}|k,l\rangle,\\
F_{3,4}|k,l\rangle & =il|k,l\rangle.\nonumber 
\end{align}
The pseudo-vacuum overlap is
\begin{equation}
\mathbf{B}=X_{1}^{L}.
\end{equation}
and the $X_{1}^{2}$ can be expressed as
\begin{equation}
X_{1}^{2}=\frac{1}{4}-\mathcal{C}_{3}.
\end{equation}
For a vector $|k,l\rangle$ the eigenvalues are
\begin{align}
\beta_{k,l} & =(k+1/2)^{L},\nonumber \\
\mathcal{F}_{k,l}^{(1)}(iu+\frac{1}{2}) & =\frac{u^{2}+\frac{1}{4}}{u^{2}}\frac{(u^{2}+\frac{(n+2)^{2}}{4})(u^{2}+l^{2})}{(u^{2}+k^{2})(u^{2}+(k+1)^{2})},\\
\mathcal{F}_{k,l}^{(2)}(iu) & =\frac{u+i/2}{u}\frac{u^{2}+(k+1/2)^{2}}{(u-i(l+\frac{1}{2}))(u-i(l-\frac{1}{2}))}.\nonumber 
\end{align}
Substituting back to the general overlap formula (\ref{eq:genOV}),
our overlap can be expressed as
\begin{align}
& \frac{\langle\mathrm{MPS}|\mathbf{u},\mathbf{v},\mathbf{w}\rangle}{\sqrt{\langle\mathbf{u},\mathbf{v},\mathbf{w}|\mathbf{u},\mathbf{v},\mathbf{w}\rangle}}=
2\sum_{k=\frac{1}{2}}^{\frac{n}{2}}\sum_{l=-k}^{k}\left(k+\frac{1}{2}\right)^{L} \times \\
& \prod_{j=1}^{N_{u}/2}\frac{(u_{j}^{2}+\frac{(n+2)^{2}}{4})(u_{j}^{2}+l^{2})}{(u_{j}^{2}+k^{2})(u_{j}^{2}+(k+1)^{2})}\sqrt{\frac{u_{j}^{2}+\frac{1}{4}}{u_{j}^{2}}}\times\\
& \prod_{j=1}^{N_{v}}\frac{v_{j}^{2}+(k+1/2)^{2}}{(v_{j}-i(l+\frac{1}{2}))(v_{j}-i(l-\frac{1}{2}))}\times\sqrt{\frac{\det G_{+}}{\det G_{-}}}.
\end{align}

\section{Conclusion}\label{conclusion}

We studied integrable matrix product states with $\mathfrak{so}_4$ symmetry  for and proved their integrability by constructing a corresponding K-matrix. Using this we established an overlap formula for the fuzzy $S^3$ MPS with the Bethe states of the $\mathfrak{so}_6$ symmetric XXX chain for arbitrary bond dimensions.  These matrix product states are related to nonsupersymmetric defect configurations of $\mathcal{N}=4$ SYM of various codimensions, and the overlap formulas compute leading order one point functions of non-protected single trace operators in the planar limit. One important question is to understand to what extend these defects make sense as quantum theories, since the classical analysis on the field theory is under less control relative to supersymmetric defects. On the other hand integrability is a much more constraining symmetry than supersymmetry, and so it is not unreasonable to expect non-supersymmetric defects to survive in the strong coupling regime. An example of this is the non-supersymmetric D3-D7 interface described by a fuzzy $S^4$ MPS in which agreement with holography were achieved in a certain double scaled limit. An important step would be to understand the holographic dual of the fuzzy $S^3$ defect, for instance be revisiting the classification of integrable boundaries for the string sigma model. One hint comes from the fact for the defect to be integrable in the complete bosonic sector of the theory, the pair structure for the Bethe roots must be compatible. The pair structure is fixed by the residual symmetry and for the fuzzy $S^3$ MPS it singles out the symmetry breaking patterns $SO(2,4)\rightarrow  SO(2,2)\times SO(2)$ and $SO(2,4)\rightarrow  SO(2)\times SO(4)$. This suggests that there is a nonsupersymmetric $D5$ configuration wrapping $AdS_3\times S^3$ which is potentially integrable which would be the holographic dual description of the fuzzy $S^3$ surface defect. Similar configurations were introduced in \cite{Georgiou:2025mgg}.

The analysis presented here hints at a systematic construction of  $\mathfrak{so}_{2n}$ symmetric integrable MPS. In general their construction relies on projections from $\mathfrak{so}_{2n+1}$ symmetric MPS which are more straightforward to construct. This projection procedure complicates the analysis since the algebraic relations between generators are somewhat obscured.  An alternative approach would be to assume some kind regularity condition on the K-matrix and so solve the KT-relation perturbatively by finding commutation relations order by order. Relatedly one might try to find all possible integrable MPS in other related systems, such as the alternating chains arising from the ABJM theory or of deformations of the XXX chain. 

\section*{Acknowledgments}

We would like to thank Marius de Leeuw for discussions. The work of A.H. is supported by ERC-2022-CoG - FAIM 101088193. T.G. is supported by the NKFIH grant PD142929 and the János Bolyai Research Scholarship of the Hungarian Academy of Science.

\bibliographystyle{elsarticle-num}
\bibliography{refs}

\end{document}